\documentclass[12pt]{iopart}
\usepackage{cite}

\newcommand{\CR}{\nonumber\\ }
\renewcommand{\d}{\mathrm{d}}
\renewcommand{\i}{\mathrm{i}}

\begin{document}
\title[Logarithmic perturbation theory for the spherical anharmonic oscillator]
{A new approach to the logarithmic perturbation theory for the
spherical anharmonic oscillator}
\author{I V Dobrovolska and R S Tutik $^\dag$}
\address{Department of Physics,
Dniepropetrovsk National University, 49050, Dniepropetrovsk,
Ukraine}

\ead{$^\dag$ tutik@ff.dsu.dp.ua}

\begin{abstract}
The explicit semiclassical treatment of the logarithmic
perturbation theory for the bound-state problem for the spherical
anharmonic oscillator is developed. Based upon the
$\hbar$-expansions and suitable quantization conditions a new
procedure for deriving perturbation expansions is offered.
Avoiding disadvantages of the standard approach, new handy
recursion formulae with the same simple form both for ground and
excited states have been obtained. As an example, the
perturbation expansions for the energy eigenvalues of the quartic
anharmonic oscillator are considered.
\end{abstract}

\submitto{\JPA} 

\maketitle

\section{Introduction}

The main task in application of the quantum mechanics is to solve
the Schr\"odinger equations with different potentials.
Unfortunately, realistic physical problems can practically never
be solved exactly. Then one has to resort to some approximations.
Most widely used among them is the perturbation theory. However,
the explicit calculation with the Rayleigh -- Schr\"odinger
perturbation theory, described in most quantum mechanics
textbooks, runs into the difficulty of the summation over all
intermediate unperturbed eigenstates. To avoid this difficulty,
alternative perturbation procedures have been proposed. They are
Sternheimer's method \cite{b1, b2}, the Dalgarno--Lewis technique
\cite{b3, b4, b5, b6}, the method developed by Zel'dovich
\cite{b7, b8, b9} and the logarithmic perturbation theory
\cite{b10, b11, b12, b13, b14, b15, b16, b17}.

It must be stressed that in both the Steinheimer method and the
Dalgarno--Lewis technique one has to solve inhomogeneus
differential equations. In solving the Schr\"odinger equation with
the Zel'dovich method, for avoiding the summation over
intermediate states the Lagrange condition of the theory of
differential equations is used.

The logarithmic perturbation theory is more straightforward in
the sense that it does not require solving any equation or
involving any additional condition. Within the framework of this
approach, the conventional way to solve a quantum-mechanical
bound-state problem consists in changing from the wave function
to its logarithmic derivative and converting the time-independent
Schr\"odinger equation into the nonlinear Riccati equation.

In the case of ground states, the consequent expansion in a small
parameter leads to handy recursion relations that permit us to
derive easily the corrections to the energy as well as to the wave
function for each order. Notice that the obtained series are
typically divergent and the evaluation of perturbation terms of
large orders is needed for applying the modern procedures of
summation of divergent series. However, when radially excited
states are considered, the standard technique of the logarithmic
perturbation theory becomes extremely cumbersome and,
practically, inapplicable for describing higher orders of
expansions.

For this reason authors of the paper \cite{b9} conclude that in
dealing with excited states the method developed by Zel'dovich has
a remarkable advantage over the logarithmic perturbation theory.
But in fact there is another approach to obtaining expansions of
the logarithmic perturbation theory that describes excited states
and the ground state exactly in the same manner by means of simple
recursion formulae.

Indeed, the above mentioned disadvantage of the standard approach
to the logarithmic perturbation theory is caused by factoring out
zeros of the unperturbed wave functions with taking into account
corrections to the positions of these nodes. On the other hand,
the number of zeros of the wave function most conveniently and
naturally is introduced in the consideration by means of
quantization conditions for the logarithmic derivative of the
wave function just as within the framework of the
Wentzel--Kramers--Brillouin (WKB) method \cite{b18, b19, b20}.
However, since the WKB-approximation is more suitable for
obtaining energy eigenvalues in the limiting case of large
quantum numbers but the perturbation theory, on the contrary,
deals with low-lying levels, the WKB quantization conditions need
change.

Recently, a new procedure based on specific quantization
conditions has been proposed to get series of the logarithmic
perturbation theory via the $\hbar$-expansion technique within the
framework of the one-dimensional Schr\"odinger equation
\cite{b21}. Avoiding disadvantage of the standard approach, this
straightforward semiclassical procedure results in new handy
recursion formulae with the same simple form both for the ground
state and excited states. Moreover, these formulae can be easily
applied to any renormalization scheme of improving the
convergence of expansions \cite{b22}.

However, in most of the practical applications of quantum
mechanics, one deals with the more complicated case involved the
three-dimensional Schr\"odinger equation with the anharmonic
oscillator potential or the screened Coulomb potential. The object
of this paper is to extend the above mentioned formalism to the
bound-state problem for the spherical anharmonic oscillator that
has numerous applications in the theory of molecules and solid
state physics. The another widely used bound-state problem for the
three-dimensional Schr\"odinger equation with the potential,
having the Coulomb singularity, will be published elsewhere.

The layout of the paper is as follows: in the next section we
summarize the main ideas of our approach concerning the
construction of new quantization conditions which are more
suitable for the semiclassical treatment of the logarithmic
perturbation theory. In section 3 the quantization conditions
obtained are used for deriving new simple recursion relations for
the calculation of perturbation expansions, which have the same
simple form both for the ground state and excited states. Section
4 demonstrates that in the case of the harmonic oscillator
potential the described approach restores the exact results for
the energy eigenvalues and eigenfunctions. Here we illustrate our
method by applying it to the example of the quartic anharmonic
oscillator, as well. The paper ends with a brief summary.

\section{The classical limit and the quantization rule}

We study the bound-state problem for a non-relativistic particle
moving in a central potential of an anharmonic oscillator admitted
bounded eigenfunctions and having in consequence a discrete
energy spectrum. This potential has a single simple minimum at
the origin and is given by a symmetric function $V(r)$ which can
be written as
\begin{equation}
V(r)=\frac{1}{2}m\omega^{2}r^{2}+\sum_{i{\geq}1}v_{i}r^{2i+2}.
\label{2}
\end{equation}
Then, by separating the angular part, the reduced radial part of
the Schr\"{o}dinger equation takes the form
\begin{equation}
 -{{\hbar}^2\over
{2m}}U''(r)+\left({\frac{{\hbar}^2l(l+1)}{2mr^2}+V(r)}\right)U(r)=EU(r).
\label{1}
\end{equation}

 Following usual practice, we apply the substitution, $ C(r)=
\hbar U'(r) / U(r) $, accepted in the logarithmic perturbation
theory and go over from the Schr\"{o}dinger equation (\ref{1}) to
the Riccati equation
\begin{equation}
\hbar C'(r)+C^{2}(r)=\frac{{\hbar}^2l(l+1)}{r^2}+2mV(r)-2m E.
\label{3}
\end{equation}

We attempt to solve it explicitly in a semiclassical manner with
series expansions in the Planck constant
\begin{equation}
E=\sum^{\infty}_{k=0}{E_{k}\hbar^{k}}\;\;\;\;\;
C(r)=\sum^{\infty}_{k=0}C_{k}(r)\hbar^{k}. \label{4}
\end{equation}
The $\hbar$-expansions under discussion simplify the problem of
taking into account the nodes of wave functions for excited
states, allowing the use of the quantization condition and the
formalism of the theory of functions of a complex variable.

In the complex plane, a number of zeros $N$ of a wave function
inside the closed contour is defined by the principle of argument
known from the analysis of complex variables. Being applied to the
logarithmic derivative, $C(r)$, it means that
\begin{equation}
\frac{1}{2\pi\i}\oint{C(r)\,\d{r}}=
\frac{1}{2\pi\i}\sum^{\infty}_{k=0}{\hbar^{k}\oint{C_{k}(r )\,\d
r}}=\hbar N. \label{5}
\end{equation}

This quantization condition is exact and is widely used for
deriving higher-order corrections to the WKB-approximation
\cite{b23,b24} and the $1/N$-expansions \cite{b25,b26,b27}. There
is, however, one important point to note. Because the radial and
orbital quantum numbers, $n$ and $l$, correspondingly, are
specific quantum notions, the quantization condition (\ref{5})
must be supplemented with a rule of achieving a classical limit
for these quantities. It is this rule that stipulates the kind of
the semiclassical approximation.

In particular, within the framework of the WKB-approach the
passage to the classical limit is implemented using the rule
\begin{equation}
\hbar\to 0\;\;\;\;n \to\infty\;\;\;\; l\to\infty\;\;\;\;\hbar
n={\rm const}\;\;\;\;\hbar l={\rm const}\label{6}
\end{equation}
whereas the $1/N$-expansion requires the condition
\cite{b25,b26,b27}
\begin{equation}
\hbar\to 0\;\;\;\; n={\rm const}\;\;\;\; l\to\infty\;\;\;\;\hbar
n\to 0\;\;\;\;\hbar l={\rm const}.\label{7}
\end{equation}

The semiclassical treatment of the logarithmic perturbation theory
proved to involve the alternative possibility:
\begin{equation}
\hbar\to 0\;\;\;\; n={\rm const}\;\;\;\;l={\rm const}\;\;\;\;
\hbar n\to 0\;\;\;\;\hbar l\to 0. \label{8}
\end{equation}

Let us consider the latter rule from  the physical point of view.
Since $\hbar l \rightarrow 0$ as $\hbar \rightarrow 0$, the
centrifugal term, $\hbar^2 l \left(l+1\right)/r^2$, has the
second order in $\hbar$ and disappears in the classical limit
that corresponds to falling a particle into the center. This
means that a particle drops into the bottom of the potential well
as $\hbar \rightarrow 0$ and its classical energy becomes $E_0 =
\min V(r) = 0$. It appears from this that the series expansion in
the Planck constant for the energy eigenvalues must now read as
$E = \sum_{k=1}^{\infty}{E_k\hbar^k}$.

Upon inserting the $\hbar$-expansions for $E$ and $C(r)$ into the
Riccati equation (\ref{3}) and collecting coefficients of equal
powers of $\hbar$, we obtain the following hierarchy of equations
\begin{eqnarray}
C_{0}^{2}(r)=2 \; m V(r)\nonumber\\
C'_{0}(r)+2 \; C_{0}(r)C_{1}(r)=-2 \; m E_{1}\nonumber\\
C'_{1}(r)+2 \; C_{0}(r)C_{2}(r)+C_{1}^{2}(r)=\frac{l(l+1)}{r^2}-2 \; m E_{2}\label{9}\\
\cdots \nonumber\\
C'_{k-1}(r)+\sum_{i=0}^{k}C_{i}(r)C_{k-i}(r)=-2 \; m
E_{k}\;\;\;k>2. \nonumber
\end{eqnarray}

In the case of ground states, the recurrence system at hand
coincides with one derived by means of the standard technique and
can be solved straightforwardly. For excited states, however, it
is necessary to take into account the nodes of the wave function,
that we intend to do by making use of the quantization condition
(\ref{5}).

It should be stressed that our approach is quite distinguished
from the WKB method not only in the rule of achieving a classical
limit but also in the choice of a contour of integration in the
complex plane. With a view to elucidating the last difference let
us now sketch out the WKB treatment of the bound-state problem
for the case of the spherical anharmonic oscillator. In the
complex plane, because the potential is described by the
symmetric function (\ref{2}), this problem has two pairs of
turning points, i.e. zeros of the classical momentum. Therefore
we have two cuts between these points: in the region $r>0$ as well
as in the region $r<0$. In spite of only one cut lies in the
physical region $r>0$, the contour of integration in the WKB
quantization condition has to encircle both cuts for the correct
result for the harmonic oscillator to be obtained \cite{b28}.

In our approach, when a particle is dropping into the bottom of
the potential well these four turning points are drawing nearer
and, at last, are joining together at the origin. Hence, all nodes
of the wave function are now removed from both positive and
negative sides of the real axis into the origin and our contour of
integration must enclose only this point and no other
singularities.

Further, let us count the multiplicity of a zero formed at $r =
0$. For the regular solution of the equation (\ref{1}), the
behaviour $r^{l+1}$ as $r \rightarrow 0 $ brings the value $l+1$.
The number of nodes of eigenfunction in the region $ r > 0$ is
equal to the radial quantum number $n$. But because the potential
(\ref{2}) is a symmetric function the same number of zeros must
be in the region $r<0$, too. Then the total number of zeros
inside the contour becomes equal to $N=2n+l+1$.

Taking into account the first order in $\hbar$ of the right-hand
side, the quantization condition (\ref{5}) is now rewritten as

\begin{equation}
\frac{1}{2\pi\i}\oint{C_1(r)\, \d r}=2n+l+1\;\;\;\;\;\;\;\;
\frac{1}{2\pi\i}\oint{C_{k}(r )\,\d r}=0 \quad \;k\neq1.
\label{10}
\end{equation}

A subsequent application of the theorem of residues to the
explicit form of functions $C_k(r)$ easily solves the problem of
the description of the radially excited states.

\section{Recursion formulae}

Let us consider the system (\ref{9}) and investigate the behaviour
of the function $C_k(r)$. From the first equation it is seen that
$C_0(r)$ can be written in the form
\begin{equation} \fl
C_0(r) = - \left[2 \, m\,  V(r) \right]^{1/2} = -m\, \omega\, r
\left( 1 + \frac{2}{m\, \omega^2} \sum_{i=1}^{\infty}{v_i
\,r^{2i}}\right)^{1/2} =\, r \sum_{i=0}^{\infty}{C_i^0 \,r^{2i}}
\label{11}
\end{equation}
where the minus sign is chosen from the boundary conditions and
coefficients $C_i^0$ are defined by parameters of the potential
through the relations
\begin{equation}
C^{0}_{0}=-m\omega\;\;\;\;\;\;\;\; C^{0}_{i}={1\over{2m\omega}}
\left({\sum_{p=1}^{i-1}{C^{0}_{p} C^{0}_{i-p}-2 m v_{i}}}\right)
\;i\geq 1.\label{12}
\end{equation}

From (\ref{11}) we recognize that $C_0(0)=0$ and, consequently,
the function $C_1(r)$ has a simple pole at the origin, while
$C_k(r)$ has a pole of order $\left(2k-1\right)$. Thus $C_k(r)$
can be represented by the Laurent series
\begin{equation}
C_{k}(r)= r^{1-
2k}\sum^{\infty}_{i=0}{C^{k}_{i}r^{2i}}\;\;\;\;\;k\geq
1.\label{13}
\end{equation}
With definition of residues, this expansion permits us to express
the quantization condition (\ref{10}) explicitly in terms of the
coefficients $C_i^k$ as
\begin{equation}
C^{k}_{k-1}= \left(2 n + l + 1 \right) \delta_{1, k}.\label{14}
\end{equation}

It is this quantization condition that makes possible the common
consideration of the ground and excited states and permits us to
derive the simple recursion formulae.

The substitution of the series (\ref{12}) and (\ref{13}) into the
system (\ref{9}) in the case $i \neq k-1$ yields the recursion
relation for obtaining the Laurent-series coefficients
\begin{equation} \fl
C^{k}_{i}=-{1\over{2C^{0}_{0}}} \left[{(3-2k+2i) C^{k-1}_{i
}+\sum_{j=1}^{k-1}\sum_{p=0}^{i} C^{j}_{p}C^{k-j}_{i-p}
+2\sum_{p=1}^{i}C^{0}_{p}C^{k}_{i-p}}-l(l+1)\delta_{2,k}\delta_{0,i}\right].\label{15}
\end{equation}

If $i=k-1$, by equating the explicit expression for $C^k_{k-1}$ to
the quantization condition (\ref{14}) we arrive at the recursion
formulae defined the perturbation corrections to the energy
eigenvalues
\begin{equation}
2mE_{k}=- C^{k-1}_{k-1}- \sum_{j=0}^{k}\sum_{p=0}^{k-1}
C^{j}_{p}C^{k-j}_{k-1-p} \; .\label{16}
\end{equation}

Thus, the problem of obtaining the energy eigenvalues and
eigenfunctions for the bound-state problem for the anharmonic
oscillator can be considered solved. The equations (\ref{15}) and
(\ref{16}) have the same simple form both for the ground and
excited states and define a useful procedure of the successive
calculation of higher orders of the logarithmic perturbation
theory.

\section{Discussion and example of application}

On applying the recursion relations obtained, the analytical
expressions for first corrections to the energy eigenvalues of
the spherical anharmonic oscillator (\ref{2}) are found to be
equal to
\begin{eqnarray}
\fl E_1  =  \frac{1 + 2\,N}{2} \,\omega \CR\fl E_2 =  \frac{\left(
            3 - 2\,L + 6\,\eta  \right) \,{v_1}}{4\,m^2\,{\omega }^2}
\CR\fl E_3  =  \frac{1 + 2\,N}{8\,m^4\,{\omega }^5} \,\left(
\left( -21 + 9\,L -
                  17\,\eta  \right) \,{{v_1}}^2 +
            m\,\left(
                15 - 6\,L +
                  10\,\eta  \right) \,{\omega }^2\,{v_2} \right)
\CR\fl  E_4 = \frac{1}{16\,m^6\,{\omega }^8}\bigg(\left(
            333 + 11\,L^2 - 3\,L\,\left( 67 + 86\,\eta  \right)  +
              3\,\eta \,\left(
                347 + 125\,\eta  \right)  \right) \,{{v_1}}^3 \CR\lo -
        6\,m\,\left(
            60 + 3\,\left( -13 + L \right) \,L + 175\,\eta  - 42\,L\,\eta  +
              55\,{\eta }^2 \right) \,{\omega }^2\,{v_1}\,{v_2}  \label{17} \\
       + m^2\,\left(
            6\,L^2 - 12\,L\,\left( 6 + 5\,\eta  \right)  +
              35\,\left(
                3 + 2\,\eta \,\left(
                    4 + \eta  \right)  \right)  \right) \,{\omega
}^4\,{v_3}\bigg) \CR\fl  E_5 = - \frac{1 + 2\,N
}{128\,m^8\,{\omega }^{11}}\bigg(\left(
                    30885 + 909\,L^2 -
                      27\,L\,\left( 613 + 330\,\eta  \right)  + \eta \,\left(
                        49927 + 10689\,\eta  \right)  \right) \,{{v_1}}^4
                        \CR\lo -
                4\,m\,\left(
                    11220 + 393\,L^2 -
                      6\,L\,\left( 1011 + 475\,\eta  \right)  + \eta \,\left(
                        16342 +
                          3129\,\eta  \right)  \right) \,{\omega \
}^2\,{{v_1}}^2\,{v_2}\CR\lo +
                16\,m^2\,\left(
                    33\,L^2 - L\,\left( 501 + 190\,\eta  \right)  +
                      63\,\left(
                        15 + \eta \,\left(
                            19 + 3\,\eta  \right)  \right)  \right) \,{\omega
}^4\,{v_1}\,{v_3} \CR\lo +
                2\,m^2\,\left(
                        3495 + 138\,L^2 + 4538\,\eta  + 786\,{\eta }^2 -
                          30\,L\,\left(
                            63 + 26\,\eta  \right)  \right) \,{\omega }^4\,{{v_2}}^2  \CR\lo-
                    4\,m^3\,\left(
                        30\,L^2 - 20\,L\,\left( 24 + 7\,\eta  \right)  +
                          63\,\left(
                            15 + 2\,\eta \,\left(
                                8 + \eta  \right)  \right)  \right) \,{\omega}^6\,{v_4} \bigg)\nonumber
\end{eqnarray}
where $N = 2 \; n+ l + 1 $, $\eta = N \left(N + 1 \right)$, $L = l
(l+1)$.

As it is seen, the obtained expansion is the expansion in powers
of the Taylor-series coefficients for the potential function.

It is also evident that for the energy eigenvalues, when $k=1$, we
readily have the oscillator approximation \cite{b29}
\begin{equation}
E_1 = \left( 2 n + l + \case{3}{2} \right) \omega.
\end{equation}

Let us demonstrate that in the case of the isotropic harmonic
oscillator our technique restores the exact solution for the wave
functions, $U(r)$, too.

Putting, for simplicity, $\hbar = m = \omega = 1$, from equations
(\ref{12}), (\ref{13}) and (\ref{15}) we find that
\begin{equation}
\label{19} C_0(r) = - r \quad\;\;\; C_k(r)=d_k r^{1-2k} \quad k >
0
\end{equation}
where coefficients $d_k$ obey the relations
\begin{eqnarray}
 d_0 & = &  -1\nonumber \\
 d_1 & = &  N\nonumber \\
 2 \; d_2 & = &  N^{2}-N-l(l+1)\label{20} \\
 \cdots \nonumber \\
 2 \; d_k & = & (3-2k)\,d_{k-1}+\sum_{j=1}^{k-1}d_j d_{k-j} \quad k>2\;.\nonumber
\end{eqnarray}

Due to the definition $C(r) = U'(r)/U(r)$, the straightforward
integration of the function $C_0(r)$ and the part of the function
$C_1(r)$ immediately gives factors $e^{-r^2/2}$ and $r^{l+1}$ in
the eigenfunctions, $U(r)$, providing their correct behaviour at
infinity and the origin. The remaining part is a polynomial
$P_{n}(r)= \sum_{k=0}^{n}p_k r^{2k}$ that satisfies the equation
\begin{equation}\label{21}
P'_{n}(r^2)/P_{n}(r^2)=\sum_{k=1}^{\infty}d_k r^{1 - 2 k} -
    (l+1)r^{-1}\;.
\end{equation}

The polynomial coefficients, $p_k$, are determined by the system
\begin{equation}\label{22}
2 \; p_m (n - m) +\sum_{j=m+1}^{n}p_j \; d_{j-m+1}=0\; .
\end{equation}
The combination of these equations, multiplied by a suitable $d_j$ with
a view to taking into account equation (\ref{20}), arrives at the
following relation between two consecutive coefficients:
\begin{equation}\label{4.406}
\frac{p_{m-1}}{p_m}=\frac{m(m+l+1/2)}{m-n-1}
\end{equation}
that is the known recursion formula for the associated Laguerre
polynomial, $L^{l+1/2}_{n}(r^2)$, of degree $n$ \cite{b30}. Thus,
the described technique restores the exact result for the
unnormalized wavefunctions of the harmonic oscillator \cite{b29},
too.

As an example of application we examine excited states as well as
the ground states of the quartic anharmonic oscillator with a
potential
\begin{equation}
V(r) = m \omega^2 r^2 /2 + \lambda r^4
\end{equation}
where $\lambda $ is a positive constant.

Equations (\ref{17}) for the energy eigenvalues now are rewritten
as
\begin{eqnarray}
\fl E_1  =  \left( \frac{1}{2} + N \right) \,\omega  \CR
\fl E_2 =
\frac{\left( 3 - 2\,L + 6\,\eta  \right) }{4\,m^2\,{\omega
}^2}\,\lambda \CR
\fl E_3 =  \frac{-\left( 1 + 2\,N \right)
\,\left( 21 - 9\,L + 17\,\eta  \right)
     }{8\,m^4\,{\omega }^5}\,{\lambda }^2 \\
\fl E_4  =  \frac{\left( 333 + 11\,L^2 - 3\,L\,\left( 67 + 86\,\eta
\right)  +
       3\,\eta \,\left( 347 + 125\,\eta  \right)  \right) }{16\,
     m^6\,{\omega }^8}\,{\lambda }^3\CR
\fl E_5  =  \frac{- \left( 1 + 2\,N \right) \,
       \left( 30885 + 909\,L^2 - 27\,L\,\left( 613 + 330\,\eta  \right)  +
         \eta \,\left( 49927 + 10689\,\eta  \right)
       \right) }{128\,m^8\,{\omega }^{11}}\, {\lambda }^4 \nonumber
\end{eqnarray}
We recall that here $N = 2 \; n+ l + 1 $, $\eta = N \left(N + 1
\right)$, $L = l (l+1)$.

It is readily seen that the use of the $\hbar$-expansion
technique does lead to the explicit perturbation expansion in
powers of the small parameter $\lambda$.

In the case of ground states, obtained expansions for the energy
eigenvalues coincide with those listed in~\cite{b31}. In the case
of excited states, our corrections coincide with corrections up to
the second order which are just only calculated in~\cite{b32}.

Thus, we derive the recursion formulae for obtaining arbitrary
order terms of the weak coupling power series for the bound-state
problem of the anharmonic oscillator. Unfortunately, as it is
known, this series is asymptotic and diverges for any finite
value of the parameter $\lambda$ that requires the use of some
procedures of improving the convergence (for references see
\cite{b33}). The most common among them are various versions of
the renormalization technique, intended to reorganize a given
series into another one with better convergence properties. In
practice, these procedures involve quite a number of the
perturbation terms and are stopped at a finite order
approximation when the change of some eigenvalue with increasing
order of approximation becomes less than the needed exactness. If
the series begins diverge before achieving the needed exactness
of the approximation, we must apply one of the methods of
summation of divergent series.

It should be noted, that the proposed formalism of the
logarithmic perturbation theory is easily adapted to the
treatment of any renormalization scheme in terms of handy
recursion relations within the framework of the united approach
\cite{b22}. Besides, the described technique can be extended to
the case of the screened Coulomb potential through the
modification of the quantization conditions, that will be
published elsewhere.

\section{Summary}

In conclusion, a new useful technique for deriving results of the
logarithmic perturbation theory has been developed. Based upon the
$\hbar$-expansions and suitable quantization conditions, new handy
recursion relations for solving the bound-state problem for a
spherical anharmonic oscillator have been obtained. These
relations can be applied to excited states exactly in the same
manner as to ground states providing, in principle, the
calculation of the perturbation corrections of large orders in
the analytic or numerical form. Besides this remarkable advantage
over the standard approach to the logarithmic perturbation
theory, our method does not imply knowledge of the exact solution
for zero approximation, which is obtained automatically. And at
last, the recursion formulae at hand, having the same simple form
both for the ground state and excited states, can be easily
adapted to applying any renormalization scheme for improving the
convergence of obtained series, as it is described in~\cite{b22}.

The another widely used case of the three-dimensional
Schr\"odinger equation with the potential having the Coulomb
singularity will be published elsewhere.

\ack This research was supported by a grant N 0103U000539 from the
Ministry of Education and Science of Ukraine which is gratefully
acknowledged.

\end{document}